\documentstyle[epsfig,12lomcon,cite]{article}

\begin{document}

\title{THE SCIENCE OF PAMELA SPACE MISSION  }

\author{ P.Picozza \footnote{e-mail: piergiorgio.picozza@roma2.infn.it },
A.Morselli \footnote{e-mail: aldo.morselli@roma2.infn.it}}

\address{University and INFN of Roma "Tor Vergata", Roma, Italy}


\maketitle\abstracts{ Many different mechanisms can contribute to antiprotons and positrons 
production, ranging from conventional reactions  up to exotic processes like 
 neutralino annihilation. The open problems are so fundamental 
 (i.e.: is the universe symmetric in matter and antimatter ?)
 that experiments in this field will probably be of the greatest 
 interest in the next years.
 Here we will summarize the present situation, 
 showing the different hypothesis and models and the experimental 
 measurements needed to lead to a more established scenario.}

\section{The main questions in Cosmic Ray Research}

There are some fundamental questions in Cosmic Ray Research that still need to be answered:

$\bullet$  Where do the particles come from?

$\bullet$  How and where do they get accelerated?

$\bullet$ How do they propagate through the interstellar medium and what kind of interactions do they encounter?

$\bullet$  What role do they play in the energy budget of the interstellar medium?

$\bullet$  Do we find hints of the existing of "exotic" particles?

The study of antimatter and antiparticle content in cosmic rays is a unique tool to investigate several physics and astrophysical phenomena. The idea of exploiting cosmic antiprotons measurements to probe
unconventional particle physics and astrophysics scenarios has a long history (see references in \cite{Wiz03})
and moved the cosmologists for several decades.  The search of antimatter is strictly connected with the baryon antibaryon asymmetry in the Universe. The present observational limit in the search of antihelium is in the order of $10^{-6}$ in the $\bar{\mathrm{He}}/\mathrm{He}$ ratio.  PAMELA will extend this limit to the  $\sim 10^{-8}$ level.

The majority of the antiproton and positron components in cosmic rays are of secondary origin, but part of them could come from other primary sources. The search and the identification of such possible sources is one of the major challenges in cosmic rays studies. In fact several observations give us a coherent picture of a Universe dominated by dark matter and dark energy: the detection of an anomalous content of antiprotons or positrons in cosmic rays could be a clear signature of a source connected with the presence of dark matter, e.g. WIMPs or Kaluza-Klein particles annihilation.  Furthermore, the precise measurement of the secondary antiproton flux is a decisive factor to well understand the propagation of cosmic rays in the galaxy, while
secondary to primary CR ratios are very sensitive  to the change of the propagation parameters. 
The knowledge of secondary antiproton flux gives complementary information with respect to that obtained from nuclei like Boron or sub-iron nuclei~\cite{jcap}.

\section{The PAMELA experiment}

\begin{figure}[p]
  \begin{center}
    \mbox{\epsfig{file= 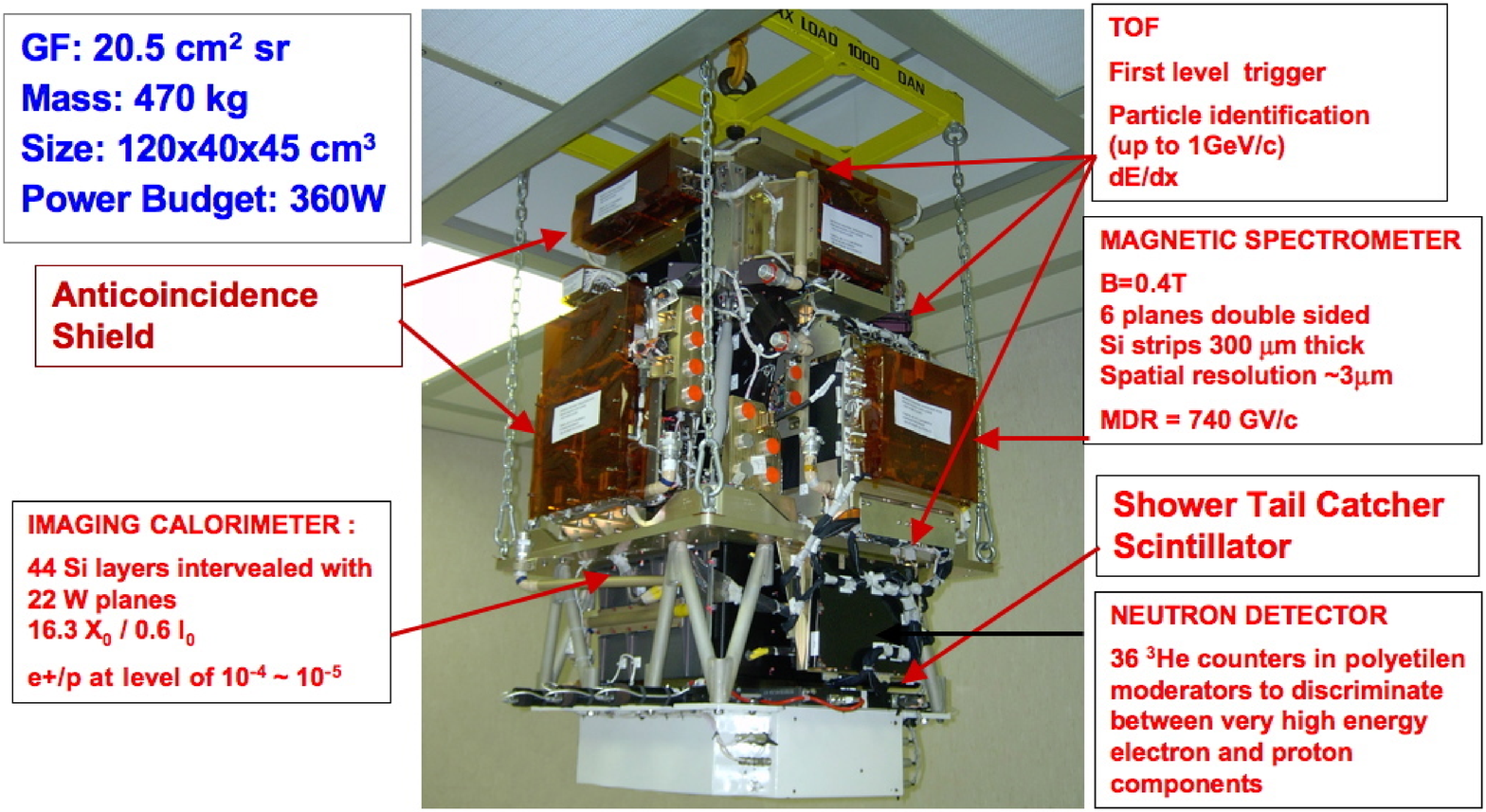,width =11.7cm} }
 \caption{the PAMELA telescope in a photo taken just prior delivery to Russia. The detector is approximately 120 cm   tall, has a mass of $\sim$ 470 kg and the power consumption is $\sim$ 360~W. \label{PAMELA}}
    \mbox{\epsfig{file= 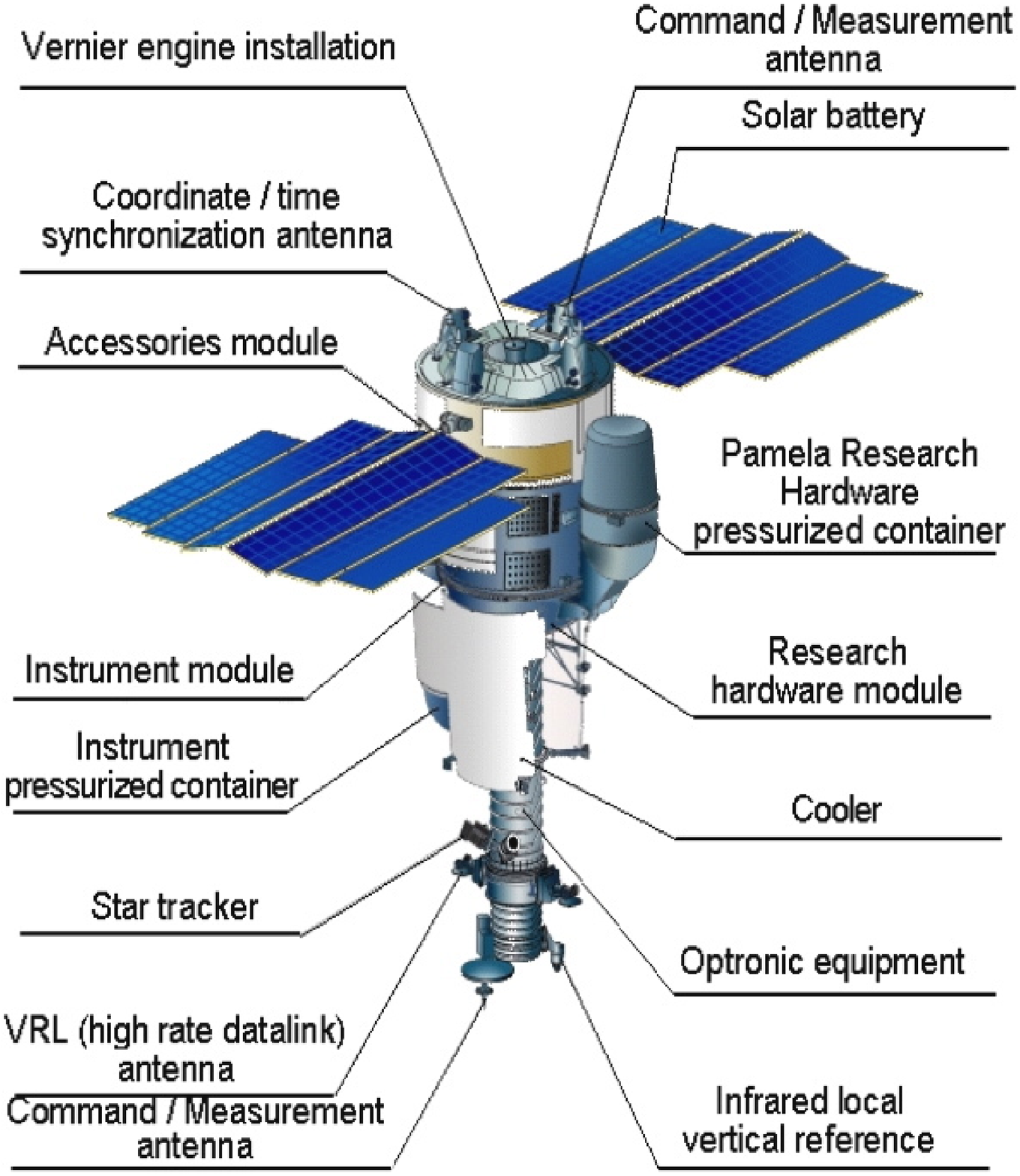,height=9.cm} }
\caption{A sketch of the Resurs  DK1 Satellite. PAMELA is mounted inside a Pressurized Container,  attached to the Satellite. \label{Resurs}}
\end{center}
\end{figure}

Figure \ref{PAMELA} shows  a picture of the apparatus that consists of: a time of flight system, a magnetic spectrometer, an anticoincidence system, an electromagnetic imaging calorimeter, a shower tail catcher scintillator and a neutron detector. PAMELA will be launched into space by a Soyuz TM2 rocket in the 2006. The position in the satellite is shown in figure~\ref{Resurs}.  The orbit is elliptical and semi-polar, with an inclination of 70$^0$ and an altitude varying between 350 km and 600 km; the mission will last at least three years. The geometric acceptance in the standard trigger configuration is $\simeq  20.5\  \mbox{cm}^2\mbox{sr}$. Moreover PAMELA is equipped with another trigger based only on the calorimeter signals: it permits to increase the acceptance of the $e^++e^-$ measurement  at high energy up to about 600 $\mbox{cm}^2\mbox{sr}$.

The main PAMELA scientific primary goals and objectives are:

$\bullet$ Search for evidence of exotic matter as heavy antinuclei  and nonbaryonic particles outside the Standard Model

$\bullet$   Understanding  formation and evolution of our Galaxy and the Universe 

$\bullet$	Exploring  the cycles of matter and 	energy in the Universe.

These objectives can be persued with  the search for structure in antiparticle spectra; the search for antinuclei and antimatter; the study of cosmic rays propagation by means of light nuclei and light isotopes spectra; the study of electron spectrum and investigation of the contribution of local sources; the study of solar physics and solar modulation.

 The capabilities of PAMELA are illustrated in table \ref{PAMELAtable}~\cite{Wiz03,Pamela}. 
\begin{table}[ht]
\begin{tabular}{|l|c|r|}
\hline 
\textbf{Particles} & \textbf{Numbers (3 years)} & \textbf{Energy range} \\ 
\hline
positrons & $\>3\times 10^5$ & 50 MeV -- 270 GeV \\  
antiprotons & $\>3\times 10^4$ & 80 MeV -- 190 GeV \\  
limit on antinuclei & &  $\sim 10^{-8}$ $\bar{\mbox{He}}/\mbox{He}$  \\
electrons & $6\times 10^6$ & 50 MeV -- 2 TeV \\  
protons & $3\times 10^8$ & 80 MeV -- 700 GeV \\
electrons+positrons &  & 50 MeV -- 10 TeV \\  
light nuclei (up to Z=6) & & 100 MeV/n -- 200 GeV/n \\
light isotopes ($^2$H, $^3$He) & & 100 MeV/n -- 1 GeV/n \\
\hline
\end{tabular}
\caption{PAMELA Capabilities}
\label{PAMELAtable}
\vspace{0.4cm}
\end{table}

\begin{figure}[ht]
  \begin{center}
  \mbox{\epsfig{file= 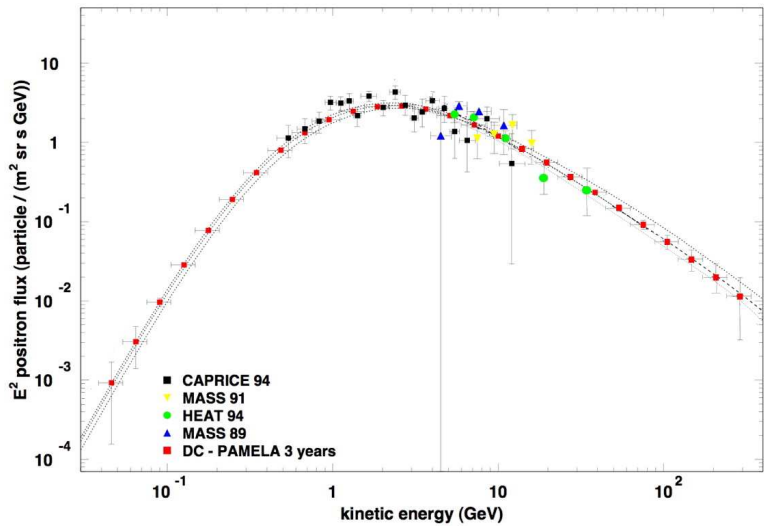, width = 10.2cm} }
 \end{center}
\caption{Experimental data (\cite{Wiz03} ) compared with PAMELA's expectations for positrons   for diffuse and convention (DC) model background \cite{jcap}. The central line corresponds to the parameters of the best B/C fit while the others show the uncertainties band.  }
\label{pamDCe}
 \end{figure}

\begin{figure}[ht]
 \begin{center}
\vspace{-0.4cm}
  \mbox{\epsfig{file= 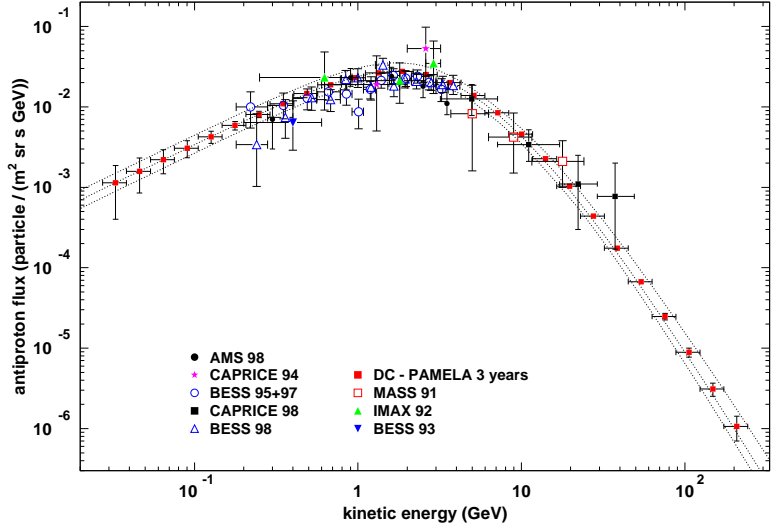, width =11cm} }
   \end{center}
\vspace{-0.4cm}
\caption{Same as figure~\ref{pamDCe} but for the antiproton flux. }
\label{pamDCp}
\end{figure}

 The expected performance of PAMELA in the case of a pure secondary positrons   and antiprotons  are shown with red  boxes, respectively in figure ~\ref{pamDCe} and ~\ref{pamDCp}
in the case of a propagation model that includes diffuse and convention~\cite{jcap}. The errors of the expected PAMELA data points include only statistical uncertainties. An average PAMELA orbit has been used to estimate the vertical geomagnetic cutoffs and the consequent expected number of antiproton and positron events.

\begin{figure}[p]
\begin{center}
  \psfig{figure=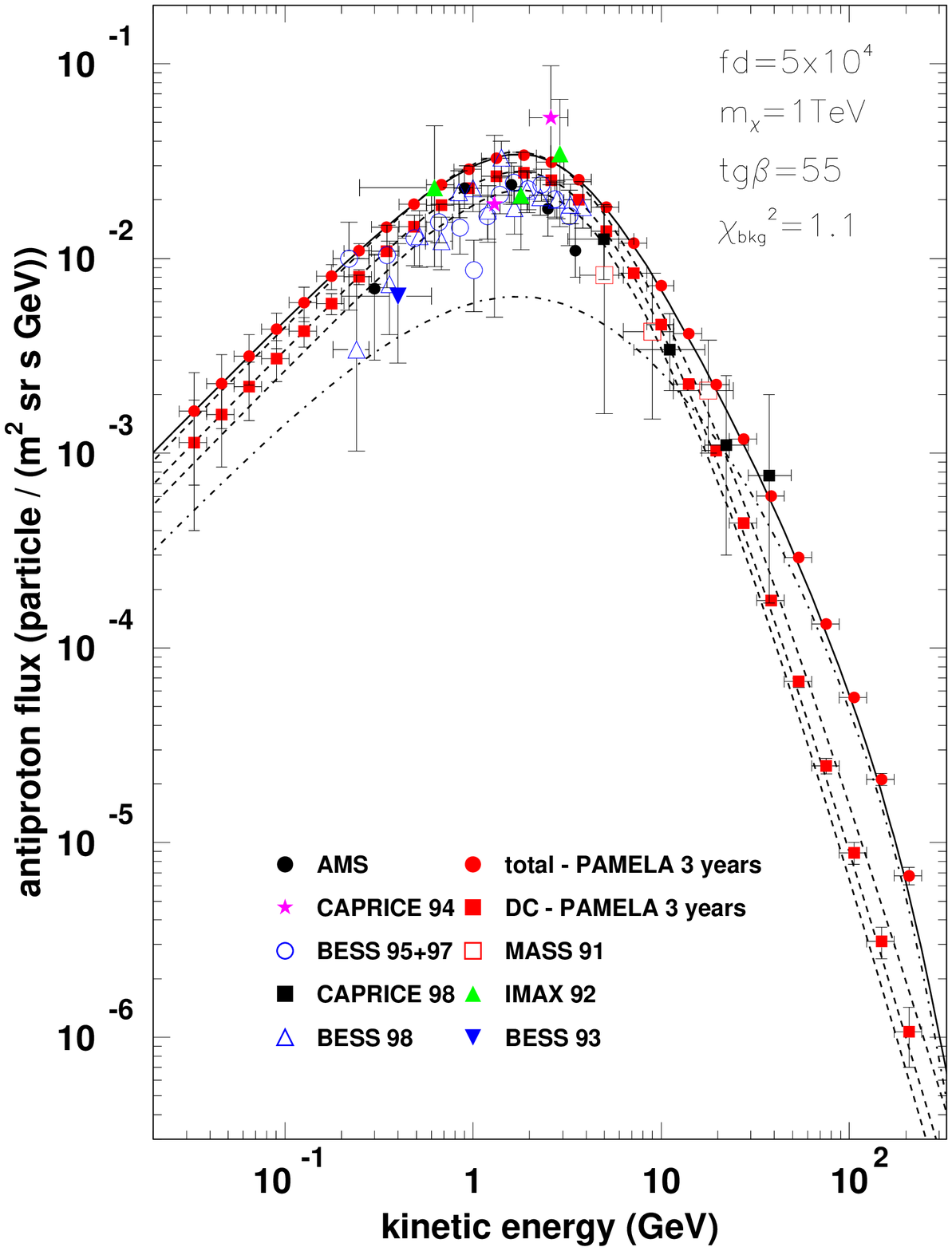,height=15.5cm}
\end{center}
\vspace{-0.9cm}
  \caption{Antiproton absolute flux:  theoretical predictions  for total uncertainty and  best B/C fit for DC model (dashed lines). Experimental data are from  \cite{Wiz03}. The PAMELA expectations points (red squares)  for DC background are for three years of data taking.  The dash-dotted line is a neutralino induced contribution for a neutralino mass of 1 TeV (see text) and a clumpiness factor $fd$ of 5 10$^4$ while the solid line is the
total contribution  calculated with the addition of the DC background and the red circles are the corresponding PAMELA points~\cite{jcap}.}
     \label{susy_antip}
\vspace{-0.4cm}
\end{figure}

The primary contribution to the $\bar{p}$ flux has been computed
using the public code DarkSUSY~\cite{ds}.  
The SUSY contribution to the $\bar{p}$
flux is shown in figure~\ref{susy_antip} for a neutralino mass of 1 TeV (obtained from a particular
choice of mSUGRA parameters) and a clumpiness factor $fd$ of 5 10$^4$. 
For different values of the  mSUGRA parameters one can  found the minimal values of the clumpiness factors $fd$ needed to disentangle a neutralino induced component in the antiproton flux with PAMELA. This factor can be computed as a function of the mSUGRA parameters. Fixing the less sensitive parameters  $A_{0}$, $\tan \beta$ and ${\rm sign} (\mu)$,  the clumpiness factor become a function of $m_{0}$ and $m_{1/2}$. 

\begin{figure}[ht]
\begin{center}
  \psfig{figure=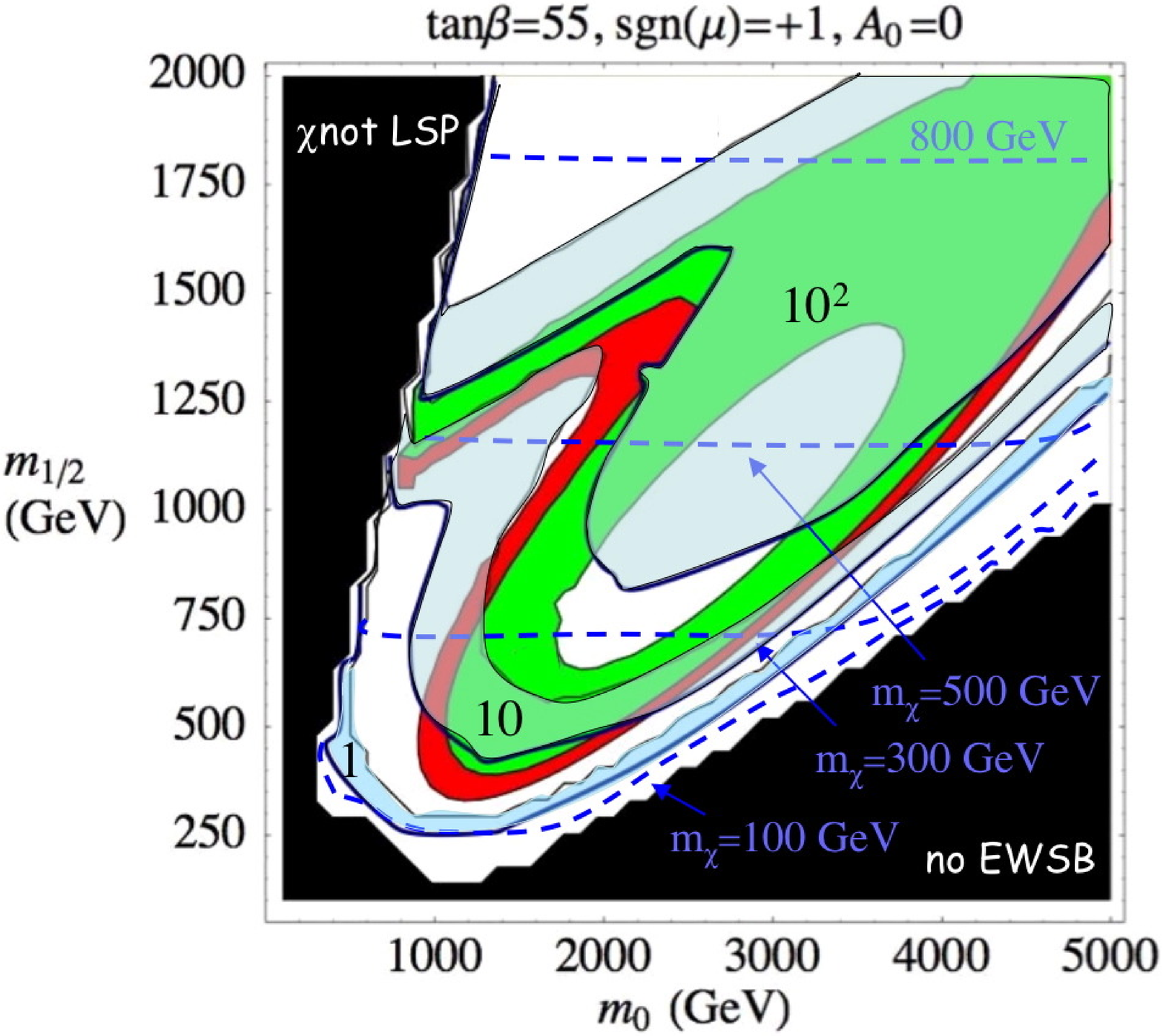,height=10.4cm}
\end{center}
\vspace{-0.4cm}
  \caption{Contour plots for the minimum $fd$ needed for a PAMELA disentanglement (upper bounds of the translucent bands) and for the maximum $fd$ allowed by current experimental data (lower bounds of the translucent bands).  The translucent regions denote the  domains that correspond to models detectable by PAMELA. .}
     \label{tg55}
\end{figure}

For each model the minimal value of the clumpiness factor $fd$ needed to satisfy both conditions has been found. As the clumpiness factor is a function of $m_{0}$ and $m_{1/2}$ parameters the contour plots can be made calculating equi-clumpiness factors lines. The results is shown in figure~\ref{susy_antip} for $\tan \beta = 55$. Black color represents the regions in the parameter space that are excluded either by accelerator bounds or because electroweak symmetry breaking is not achieved or because the neutralino is not the lightest supersymmetric particle. Red (dark shaded) are domains with $\Omega h^{2}$ in the WMAP region $0.09 < \Omega h^{2} < 0.13$, while green (light shaded) are the parameter space domains with $0.13 < \Omega h^{2} < 0.3$.  The equi-neutralino mass contours is in blue dashed lines~\cite{jcap}. Other example are in~\cite{jcap}. It can be see that PAMELA will be able to disentangle a neutralino induced component for halo models that has  $fd $ as low as $\sim 10$.

\begin{figure}[ht]
\begin{center}
  \psfig{figure=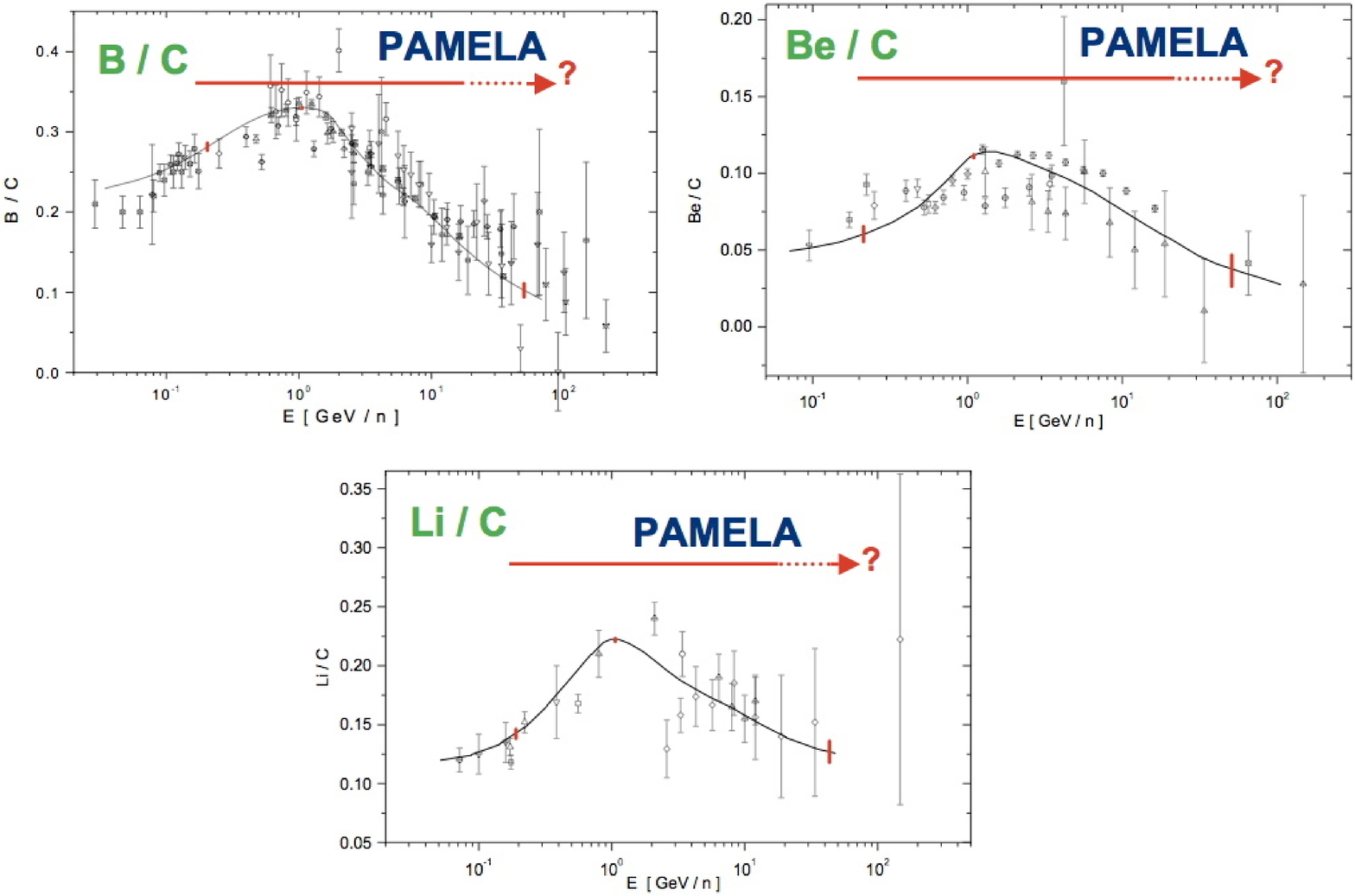,height=8.0cm}
\end{center}
  \caption{Present status of secondary to primary ratio measurements with the
 PAMELA energy range capability and  some example of the PAMELA statistical errors.}
     \label{Sec_Prim_ratio}
\end{figure}

PAMELA will give an fundamental contribution to the definition of the propagation parameters
with the measure of different cosmic ray ratio and of Hydrogen and Helium spectra.   Some examples  are shown  
 in figure~\ref{Sec_Prim_ratio}.
 
 Solar physics is another relevant items that PAMELA will deeply study.
 In particular PAMELA will measure: 
 \begin{itemize}
\item  Solar Modulation effects
 
 \item  High energy component of electron and positron in Solar Proton Events (from 50 MeV to 100 GeV)
 
 \item Nuclear composition of gradual and impulsive events
 
 \item  $^3$He and $^4$He isotopic composition.
 \end{itemize}

There are osservational evidence that the negative charge component of galactic cosmic rays is modulated in the heliosphere differently than the positive one. The modification and modulation of Galactic Cosmic Ray spectra in the heliosphere complicate the interpretation of the exotic matter results at low energy. PAMELA will monitor the solar modulation for a significative part of the solar cycle.

\section{Conclusion}

PAMELA is the first space experiment which will measure the antiprotons and positrons to high energies, $\sim 200 $ GeV, with an unprecedented statistical precision.
It will set a new lower limit of finding antihelium. It will look for dark matter candidates. It will provide measurements on elemental spectra and low mass isotopes with an unprecedented statistical precision and will help to improve the understanding of particle propagation in the interstellar medium. It will be able to measure the high energy tail of solar spectra and  for the first time solar positrons. It will be able to measure electrons at very high energy to discover sources near the solar system.

\end{document}